\documentclass[10pt,a4paper,twoside]{article}
\usepackage{epsfig}
\usepackage{baltlat6}
\usepackage{array}
\usepackage{here}
\usepackage{subfigure}
\begin{document}
\ \
\vspace{0.5mm}
\setcounter{page}{277}
\vspace{8mm}

\titlehead{Baltic Astronomy, vol.\,20, 558-565, 2011}

\titleb{ANALYSIS OF HYDROGEN CYANIDE HYPERFINE SPECTRAL COMPONENTS TOWARDS STAR FORMING CORES}

\begin{authorl}
\authorb{R. M. Loughnane}{1}, 
\authorb{M. P. Redman}{1},
\authorb{E. R. Keto}{2},
\authorb{N. Lo}{3,4,5},
\authorb{M. R. Cunningham}{4},
\end{authorl}

\begin{addressl}
\addressb{1}{Centre for Astronomy, School of Physics, National University of Ireland, \\  
University Road, Newcastle, Galway, Ireland; \\
loughnane.robert@gmail.com, matt.redman@nuigalway.ie}
\addressb{2}{Havard Smithsonian Center for Astrophysics,\\
60 Garden Street, Cambridge, MA 02138; keto@harvard.edu}
\addressb{3}{Departamento de Astronom\'{i}a, Universidad de Chile, Camino El Observatorio 1515, Las Condes,
Santiago, Casilla 36-D, Chile (current address)}
\addressb{4}{School of Physics, University of New South Wales, Sydney, NSW 2052, Australia}
\addressb{5}{Laboratoire AIM Paris-Saclay, CEA/Irfu - Uni. Paris Did\'{e}rot - CNRS/INSU, 91191 Gif-sur-Yvette, 
France}
\end{addressl}

\submitb{Received: 2011 June 20; accepted:}

\begin{summary} 
Although hydrogen cyanide has become quite a common molecular tracing species for a variety of astrophysical sources, it, however, exhibits dramatic non-LTE behaviour in its hyperfine line structure. Individual hyperfine components can be strongly boosted or suppressed. If these so-called hyperfine line anomalies are present in the HCN rotational spectra towards low or high mass cores, this will affect the interpretation of various physical properties such as the line opacity, $\rm\tau_{\lambda}$, and excitation temperature, $\rm T_{ex}$, in the case of low mass objects and infall velocities, $\rm v_{in}$, in the case of their higher mass counterparts. This is as a consequence of the direct affects that anomalies have on the underlying line shape, be it with the line structural width or through the inferred line strength. This work involves the first observational investigation of these anomalies in two HCN rotational transitions, J=1$\rightarrow$0 and J=3$\rightarrow$2 towards both low mass starless cores and high mass protostellar objects. The degree of anomaly in these two rotational transitions is considered by computing the ratios of neighboring hyperfine lines in individual spectra. Results indicate some degree of anomaly is present in all cores considered in our survey, the most likely cause being line overlap effects among hyperfine components in higher rotational transitions.
\end{summary}


\resthead{Analysis of HCN hyperfine spectral components towards star forming cores}
{R. Loughnane, M. Redman, E. Keto, N. Lo, M. Cunningham}

\sectionb{1}{INTRODUCTION}

To be a good probing species for dynamical conditions that lead to the onset of star formation, a molecule must be abundant enough so that it can be detected as well as excitable in higher density regions. The purpose of the latter is such that it can be used to trace the deep dense inner regions rather than the sufficiently lower density outer regions. Being a nitrogen-bearing species, HCN fulfills these properties and having first been discovered in the cosmos by Snyder \& Buhl (1979), it has since been used routinely as an infall tracer in low mass starless cores (Tafalla et al. 2006; Sohn et al. 2007) and also in high mass star formation regions (Wu et al. 2005).

HCN has a hyperfine structure due to the nuclear quadrupole moment of the $\rm^{14}N$. This extends the use of a particular species since the optical depth and self-absorption could be measured by examining individual hyperfine lines while a measure of the excitation temperature may be gleaned from an examination of different rotational transitions.

\begin{figure}[h!]
\vbox{
\centerline{\psfig{figure=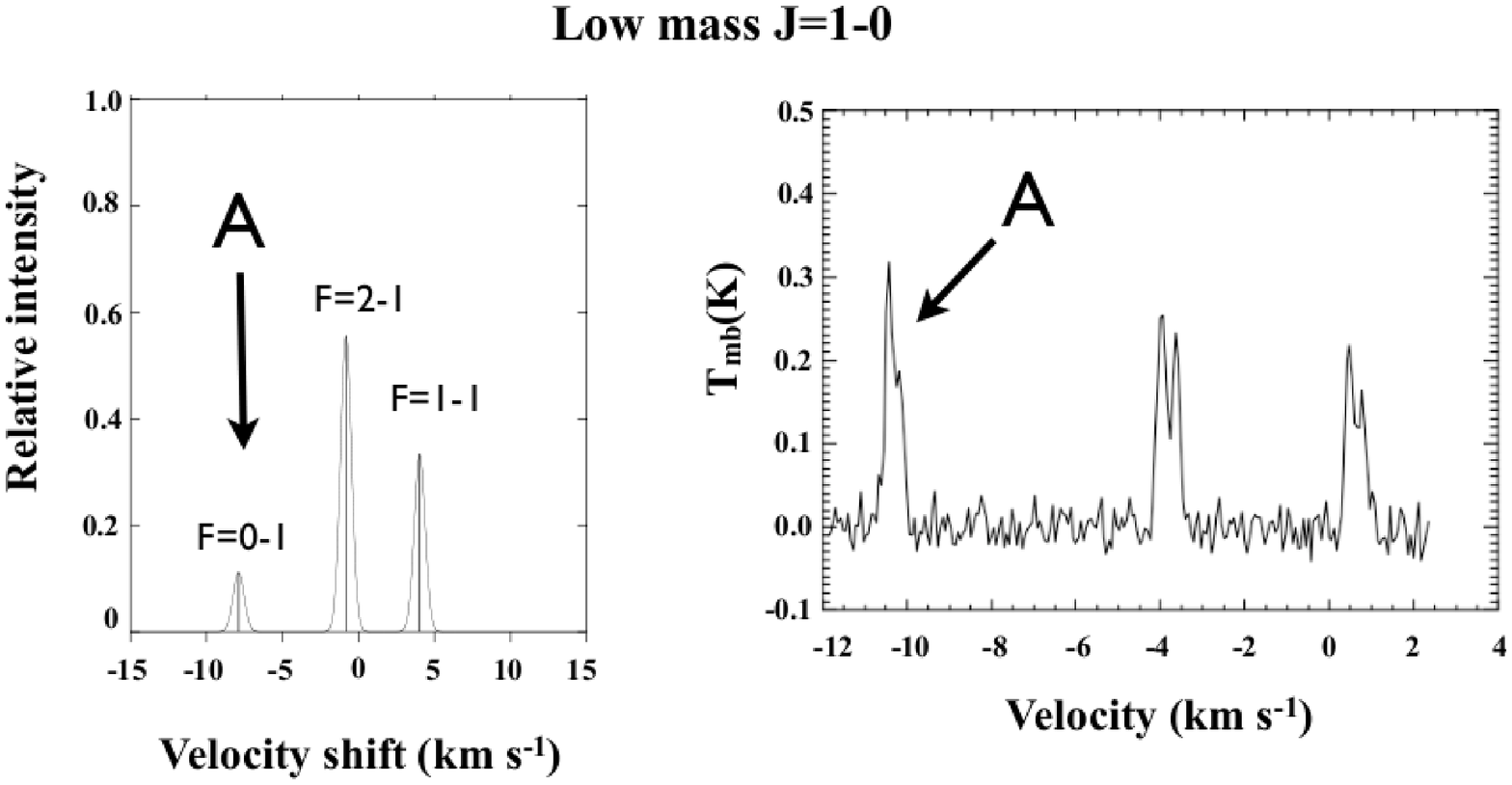,height=150pt,angle=0,clip=}}
\centerline{\psfig{figure=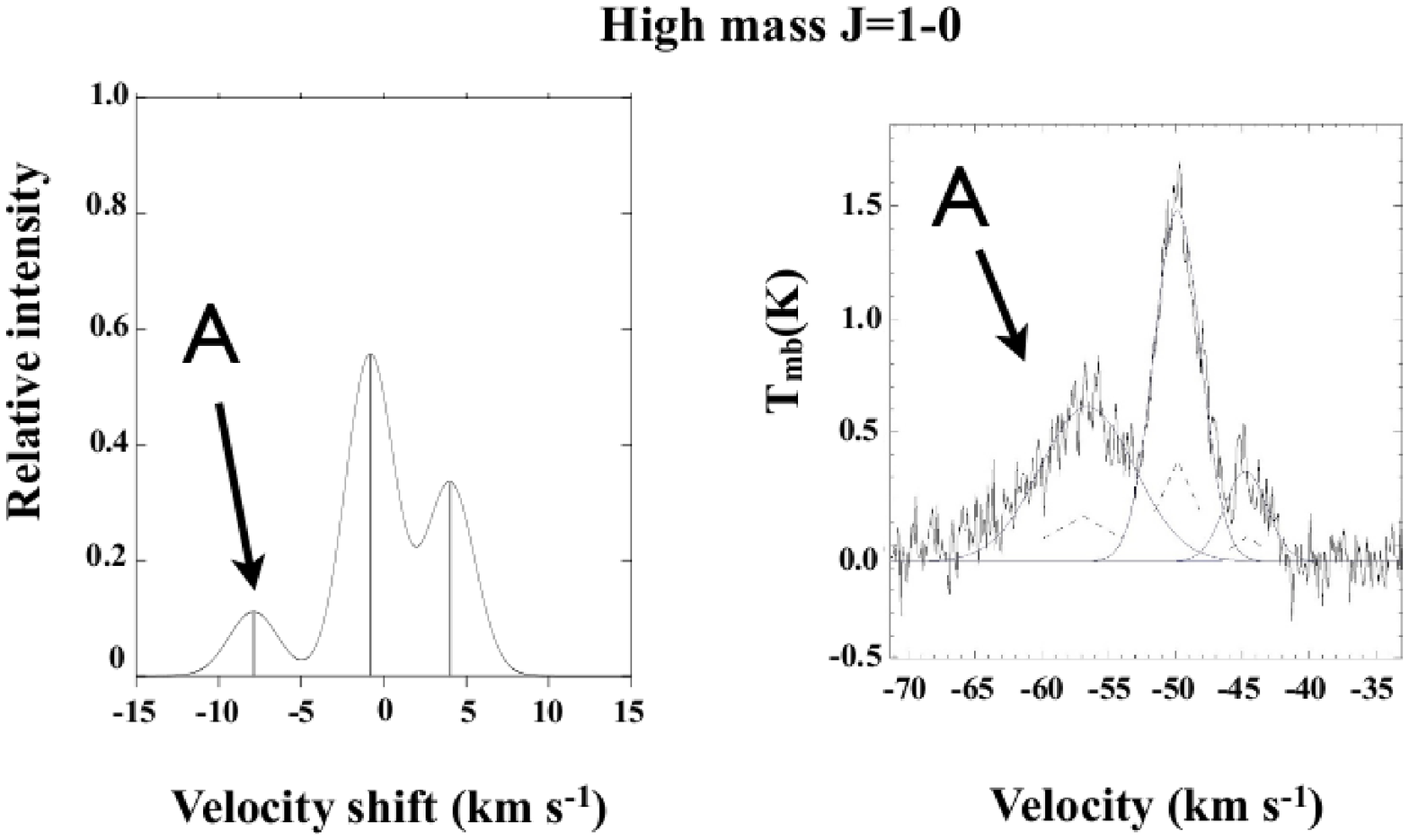,height=150pt,angle=0,clip=}}}
\vspace{1mm}
\captionb{1}
{Top (\textit{left}): Optically thin HCN J=1$\rightarrow$0 for a low mass quiescent cloud core. Top (\textit{right}): TRAO observation of low mass protostellar core, L1197. Bottom (\textit{left}): Optically thin HCN J=1$\rightarrow$0 for high mass turbulent cloud core. Bottom (\textit{right}): Mopra observation of high mass core in the G333-6.02 molecular cloud.}
\label{lowmasshighmass10}
\end{figure}

Firstly identified by Gottlieb et al. (1975), the hyperfine anomalies present in HCN J=1$\rm\rightarrow$0 were confirmed towards the low mass starless core TMC-1 by Walmsley et al. (1982) who demonstrated that the hyperfine components in the core spectrum were present in ratios that were not in thermal equilibrium with one another. Such anomalies are seen to be present towards mostly nitrogen-bearing species, i.e. molecules  possessing the $\rm^{14}N$-isotope, where the individual hyperfine lines can be seen to be boosted or suppressed far beyond LTE or saturation conditions.

\sectionb{2}{HCN AS AN OBSERVING TOOL}

The lower rotational transitions of HCN are excellent tracers of dense molecular gas in star-forming clouds. The relation $\rm n_{crit}\propto\mu^2\nu_{J\rightarrow J-1}^2$ (Papadopolous 2007) for critical densities of rotational transitions in which $\nu_{J\rightarrow J-1}$  is the frequency of an optically thin line, allow HCN transition lines to trace approximately 100-500 times denser gas than corresponding CO transitions.

\begin{figure} 
\centerline{\psfig{figure=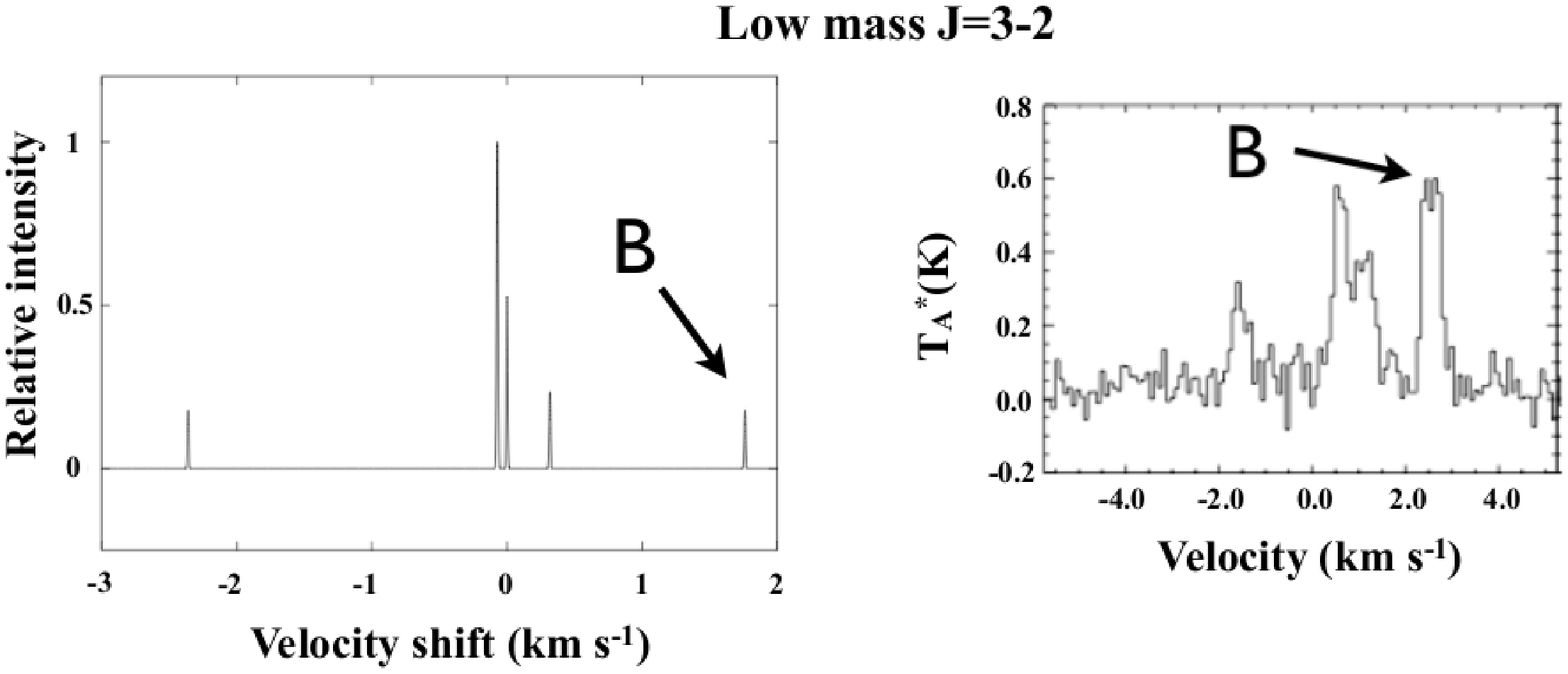,height=130pt,angle=0,clip=}}
\captionb{2}
{\textit{Left:} Optically thin HCN J=3$\rightarrow$2 compared with \textit{Right:} a JCMT observation of L1622A-2.}
\label{lowmass32}
\end{figure}

HCN is less prone to freeze-out compared to other abundant $\rm H_2$ mass tracers such as HC$\rm O^+$ or C$\rm ^{17}$O (Freed \& Mangum 2005), and therefore remains abundant in the gas phase in the cold central regions of star-forming cores. The hyperfine structure of HCN is due solely to the non-vanishing electric quadrupole moment of the end $\rm^{14}N$ nucleus, leading to line splitting of several MHz, resulting in an easily identified split line structure. Hyperfine line ratios prove an indispensable tool in the constraining of optical depth estimations of a given region in the cloud.

As mentioned in the foregoing section, anomalies scupper the usefulness of HCN in tracing the dynamics of interstellar regions. This is due principally to the variation in relative intensity among hyperfine components. Under LTE or optically thin conditions, the relative weightings of the hyperfine components in the lower rotational transition are of the form 1:5:3. For HCN J=3$\rightarrow$2, four of the six hyperfine lines are not spectrally resolved and the spectrum takes the appearance of one strong central component and two satellite lines, with a ratio of 1:25:1 between them. 

In Figure 1, the line marked ``A'' (JF=10$\rightarrow$01) in the low-mass spectrum is boosted far beyond its expected strength. As for the high-mass spectrum in the bottom panel, given that the hyperfine components are broader on account of the increased turbulence, the component marked ``A'' is again seen to be boosted as well as broadened relative to the other components. This strongly suggests that line overlap effects at higher energy levels are significant. For our analysis of the HCN J=3$\rightarrow$2 hyperfine spectrum, we defined the leftward component as the $\rm\Delta F$=$0^{-}$ hyperfine branch, the rightward component as the $\rm\Delta F$=$0^{+}$ hyperfine branch and the central four components that remain unresolved towards even the most quiescent low mass cores as the $\rm\Delta F$=1 hyperfine branch. This nomenclature is in keeping with the relative changes in hyperfine quantum number for the components involved (except the JF=32$\rightarrow$23, where $\rm\Delta F$=-1, which is part of the central group of 4 components) as well as the relative velocity shifts of the two $\rm\Delta F$=0 components, (-)-leftward versus (+)-rightward. In relation to Figure 2, the 1:25:1 proportion of intensity amongst the hyperfine branches in optically thin conditions is seen to be unmatched towards the low-mass source with the component marked ``B'' (JF=33$\rightarrow$23) appearing strongly anomalous.

\sectionb{3}{OBSERVATIONAL DATA}

In order to quantify the presence of HCN hyperfine anomalies in low mass starless cores, observations of both the HCN J=1$\rightarrow$0 and J=3$\rightarrow$2 lines for a selection of low-mass starless sources, mainly towards the Tuarus-Auriga and Ophiuchus molecular clouds, were collected. The sources were primarily chosen from a comprehensive HCN J=1$\rightarrow$0 survey in HCN and HNC, catalogued by Sohn et al. (2007) using the Taeduk Radio Astronomical Observatory (TRAO). The low-mass HCN data forming part of this work is comprised of both the TRAO data observed by Sohn et al. (2007) of 65 sources observed in J=1$\rightarrow$0 and 29 of these same sources observed in J=3$\rightarrow$2. 

The cores observed in J=3$\rightarrow$2 were selected on the basis of how bright the central F=2$\rightarrow$1 hyperfine component is in the HCN J=1$\rightarrow$0 rotational transition since it was anticipated that if this particular hyperfine transition was strong then the probability for a detection in the J=3$\rightarrow$2 rotational transition of HCN was improved. Cores that were also clearly anomalous in the HCN J=1$\rightarrow$0 line were further prioritized. These observations were carried out at the James Clerk Maxwell Telescope (JCMT) between September 2007 and July 2008 in band 5 ($\tau_{225}\ge0.2$) conditions. Single-point observations were obtained for 30 sources in position-switching mode with a predefined off-position of 50$\rm0^{\prime\prime}$ in an arbitrary direction. The ACSIS digital autocorrelation spectrometer with a bandwidth of 250-MHz was used providing a velocity resolution of $\sim0$.034km$\rm s^{-1}$. The receiver noise temperature (DSB mode) was 510-850K. The telescope main-beam efficiency was 0.69 and the half-power beamwidth (HPBW) was $\sim$2$0^{\prime\prime}$ in the range 225-285GHz. The reduced J=3$\rightarrow$2 data of each low-mass source along with its corresponding J=1$\rightarrow$0 spectrum will be published in Loughnane et al. (2011).

\begin{figure}
\centering
\mbox{\subfigure{\includegraphics[width=2in]{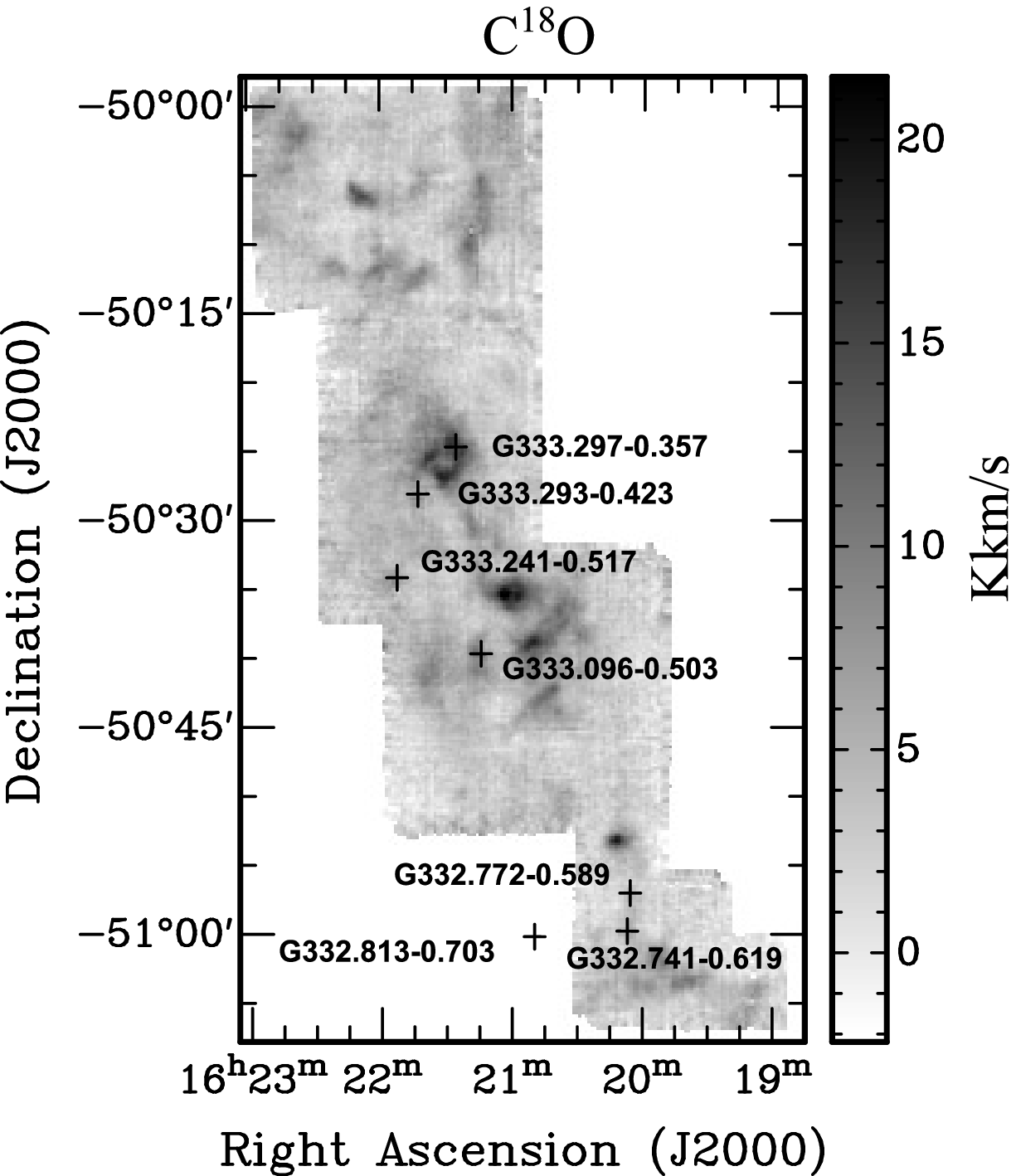}}\quad
\subfigure{\includegraphics[width=2in]{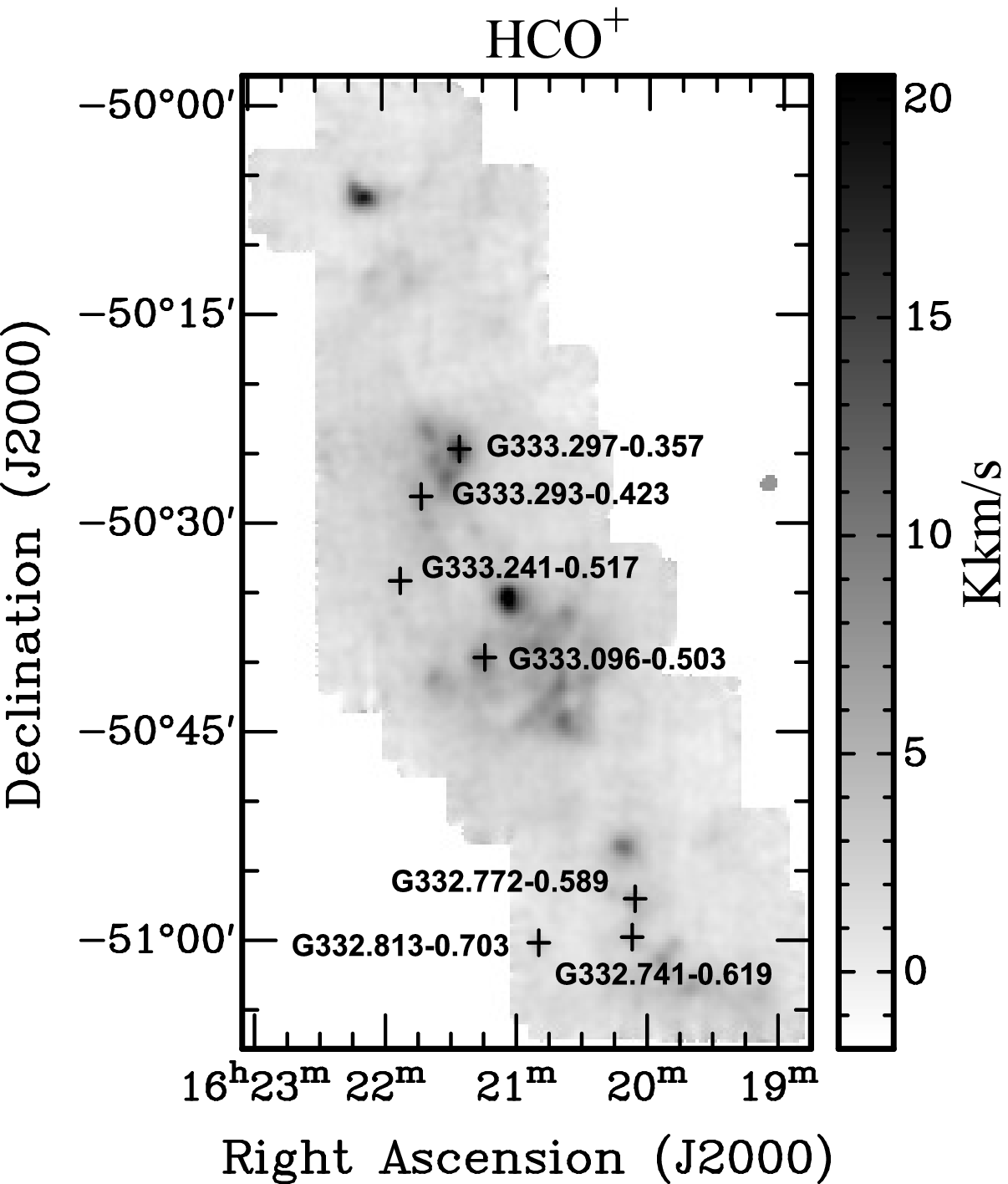} }}
\captionb{3}{C$\rm^{18}O$ J=1$\rightarrow$0 ($\textit{left}$) and HC$\rm O^+$ J=1$\rightarrow$0 ($\textit{right}$) integrated intensity (zeroth-moment) maps of the G333 giant molecular cloud. The crosses mark the positions of the seven anomalous HCN emission sources presented in this work.} \label{fig12}
\end{figure}

For the high mass HCN J=1$\rightarrow$0 data, individual sources in the G333 molecular cloud associated with the RCW 106 H$\textsc{II}$ region were observed (Lo et al. 2009). This giant molecular cloud complex spans an $\sim0$.7 de$\rm g^2$ region and is $\sim3$.6~kpc away. The data analysed as part of this work was collected with the 8-GHz wide band Mopra spectrometer (MOPS) centred at 87-GHz. The FWHM beamsize is $\sim36^{\prime\prime}$ (Ladd et al. 2005). Full details of the observational setup is available in Lo et al. (2009). We used HNC J=1$\rightarrow$0 data to locate single sources of emission in this data since this species is co-spatial to HCN but is less abundant. Maximal positions of intensity in the HCN J=1$\rightarrow$0 line that correspond to a single peak in the HNC J=1$\rightarrow$0 line (actually exhibits an unresolved hyperfine structure) represent our seven young massive protostellar cores as indicated in the integrated intensity zeroth moment maps in Figure 3. The reduced core spectra for these seven sources are displayed in Figure 5 of Loughnane et al. (2011).

\sectionb{4}{RESULTS \& CONCLUSIONS}

From the high mass HCN J=1$\rightarrow$0 spectrum in Figure 2 (Mopra data), it is clear that the three hyperfine components are merged to a degree. In order to analyse the degree of anomaly in these high mass sources, it is therefore necessary to come up with a method of dealing with the integrated emission that falls in each overlapped region. Therefore for each high mass spectrum, the individual components were approximated as gaussian profiles with associated function given by Equation~\ref{eq:1} below

\begin{equation}
y(x) = A\exp{\left(-0.5\left[\frac{v-c}{\sigma}\right]^2\right)} \label{eq:1}
\end{equation}

Rearranging Equation~\ref{eq:1}, we were able to compute the velocity $\rm v_{mid}$, defined as the velocity position between two hyperfine components where the height (or strength) of the integrated emission in the common region is at its greatest. The velocity, $\rm v_{mid}$, is the solution to the quadratic equation given by Equation~\ref{eq:2} below.  

\begin{equation}
(\sigma_2^2-\sigma_1^2)v^2+2(\sigma_1^2c_2-\sigma_2^2c_1)v
-\left[2\sigma_1^2\sigma_2^2\ln\left(\frac{A_1}{A_2}\right)+
\sigma_1^2c_2^2-\sigma_2^2c_1^2\right] = 0 \label{eq:2}
\end{equation}

Using this velocity and a calculation of the integrated emission in each of the estimated gaussian profiles with similar linewidths and line strengths as the actual observed hyperfine components, the common intensity regions between two high mass hyperfine lines could be calculated. The details of this calculation are left to the appendices of Loughnane et al. (2011). The common intensity in each case was apportioned to each of the overlapping hyperfine lines by way of their non-overlapping intensities. In this way, an integrated emission calculation was computed for each hyperfine line belonging to an individual high mass core.

\begin{figure*}[!Ht]
\begin{center}
\subfigure{%
\label{fig:lowmassrat}\includegraphics[scale=0.28]{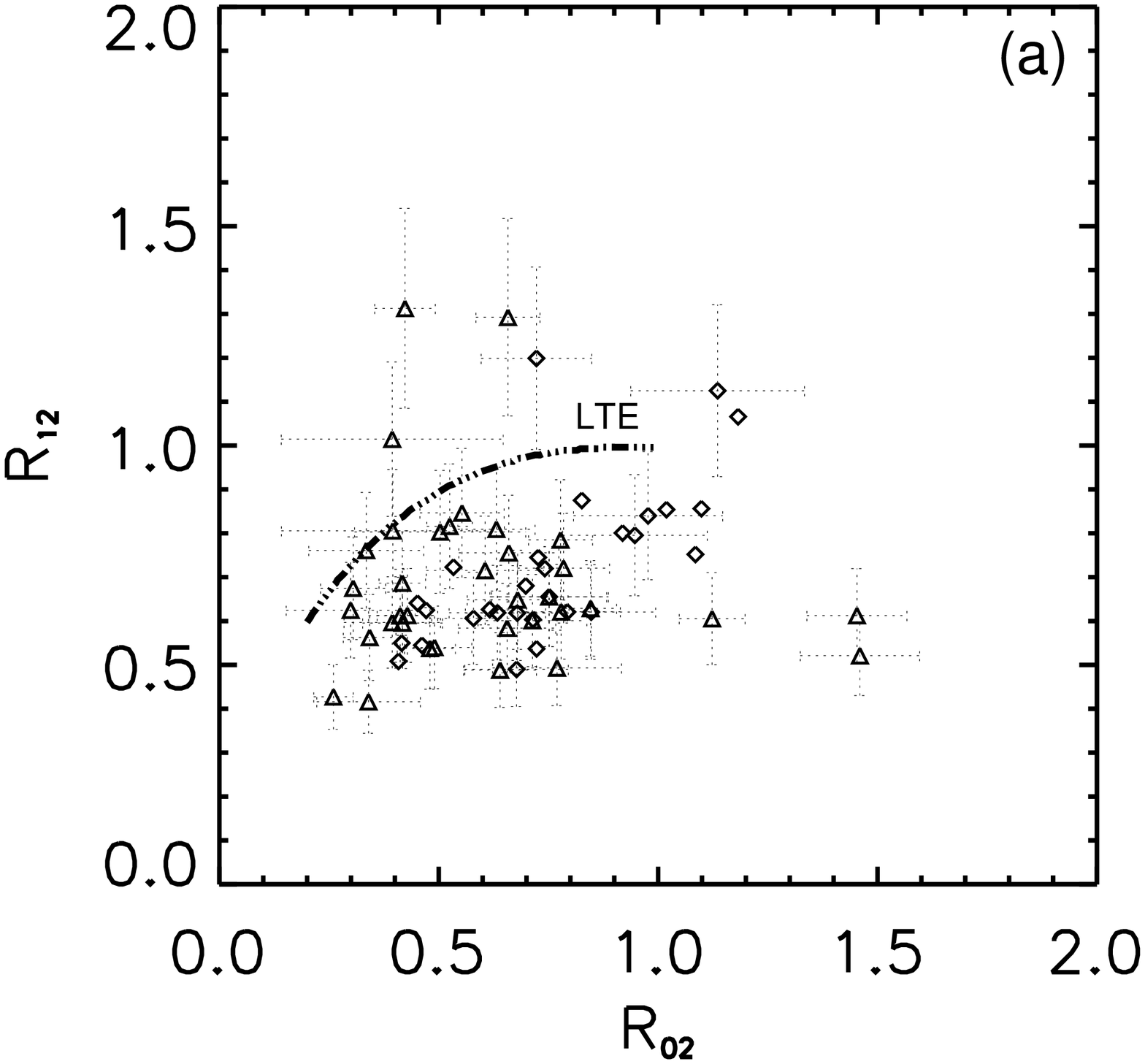}}
\hspace{0.5cm}
\subfigure{%
\label{fig:highmassrat}\includegraphics[scale=0.28]{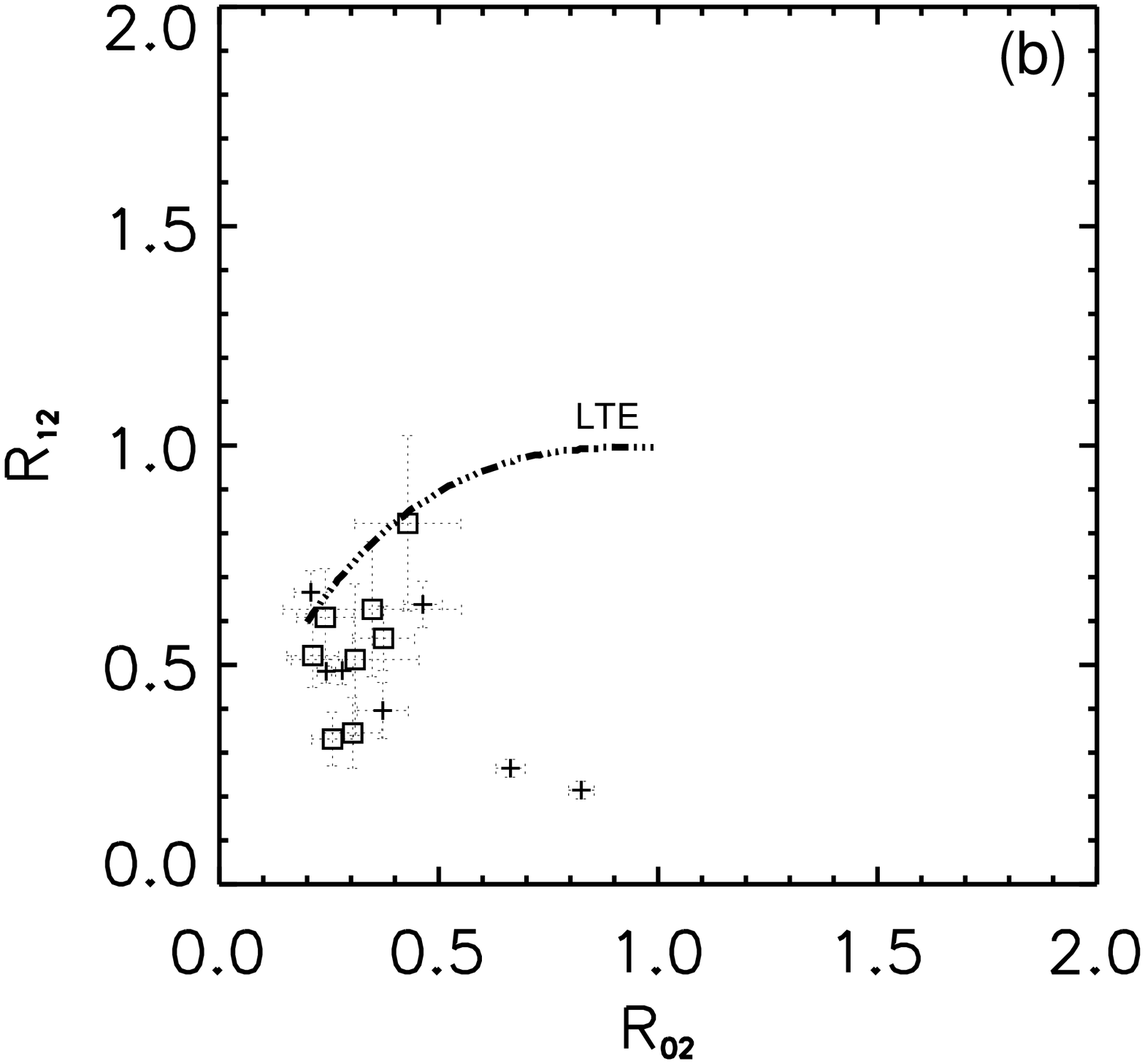}}
\subfigure{%
\label{fig:compmassrat}\includegraphics[scale=0.28]{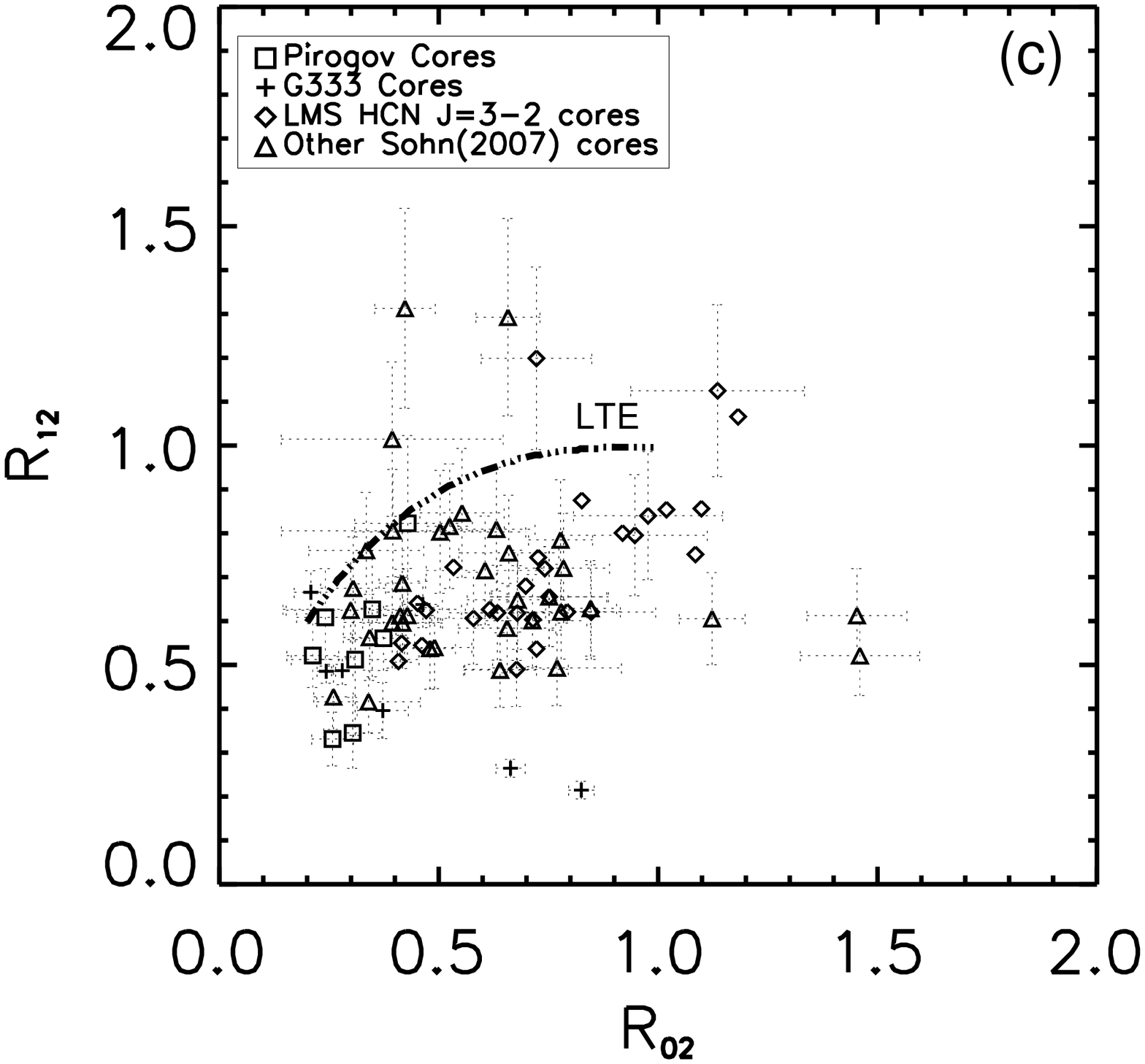}}
\end{center}
\captionb{4}
{HCN J=1$\rightarrow$0 hyperfine ratios, $\rm R_{12}$ plotted against $\rm R_{02}$, for (a) 64 low-mass starless sources located in the Taurus-Auriga and Ophiuchus molecular clouds and (b) 15 high-mass objects considered in this work, made up from seven high-mass protostellar cores located in G333-6.02 and the eight sources in figure 1 of Pirogov (1999). In (c), both sets of object's ratios are plotted in tandem. In each of the three panels, the dot-dashed curve marked LTE represents the line along which cores displaying HCN hyperfine spectral components obeying the Boltzmann equations of thermodynamic equilibrium would lie, in that part of the cloud; hence local-thermodynamic equilibrium (LTE).}
\label{fig:ratiosgraph}

\end{figure*}

For each of the seven sources observed in the G333 molecular cloud, we computed the relative intensity ratios $\rm R_{02}$; the ratio of the intensity of the F=0$\rightarrow$1 hyperfine component to the ratio of the intensity of the F=2$\rightarrow$1 hyperfine component, and $\rm R_{12}$; the ratio of the intensity of the F=1$\rightarrow$1 component to the ratio of the intensity of the F=2$\rightarrow$1 hyperfine component. These sets of intensities for all seven sources were plotted against one another along with the same ratios for eight sources considered as part of Pirogov (1999). The resulting plot is displayed in the top right-hand panel of Figure 4 (Fig. 4(b)) below. The same analysis was carried out for 65 of the low-mass sources considered in Sohn et al. (2007), with the resulting plot displayed in the top left-hand panel of Figure 4 (Fig. 4(a)) below. Both sets of data points are included in tandem in the bottom panel (Fig. 4(c)).  

What this analysis demonstrates is that high mass cores are more likely to exhibit intensity ratios close to the 1:5:3 formation common to optically thin regions. In Figure 4(b), the high mass cores are more tightly packed than the low mass objects in Figure 4(a), which show degrees of anomaly that are quite varied. In each of these panels, the dashed curve represents the progression of LTE ratios for different optical thicknesses.

\begin{figure}[h]
\vbox{
\centerline{\psfig{figure=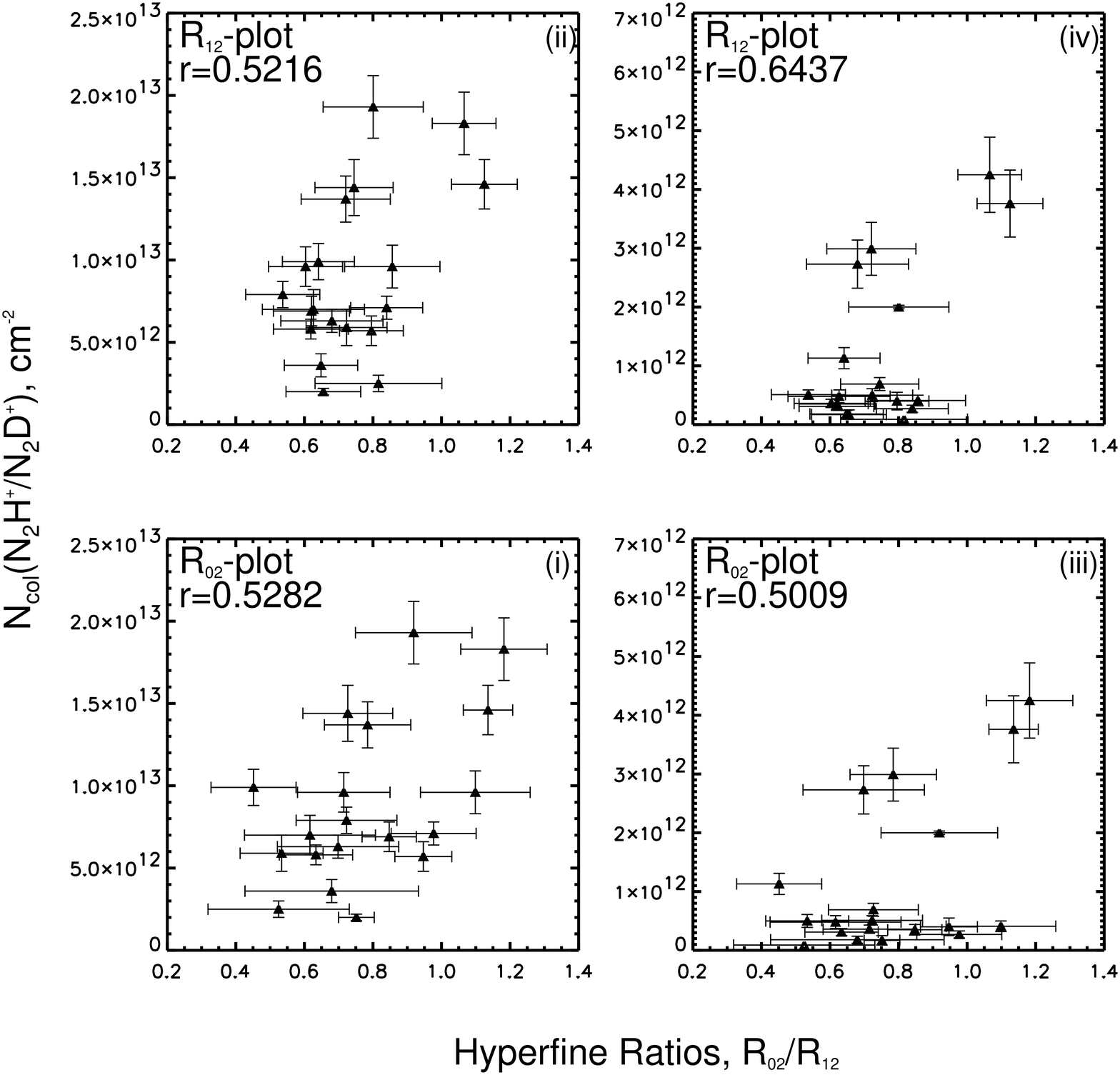,width=110mm,angle=0,clip=}}
\vspace{1mm}
\captionb{5}
{Hyperfine ratios $\rm R_{12}$ ($\rm\textit{top}$) and $\rm R_{02}$ ($\rm\textit{bottom}$) versus $\rm N_2H^+$ and $\rm N_2D^+$ column densities (both J=1$\rightarrow$0 transition). The correlation coefficient r is recorded in the top left corner.}
}
\end{figure}

Taking the sets of ratios computed above, individually, we set about comparing the degree of anomaly for low-mass starless cores with a physical parameter of the core. The columnar densities of both the $\rm N_2H^+$ and $\rm N_2D^+$ J=1$\rightarrow$0 transitions were taken from Crapsi et al. (2005) and plotted against the hyperfine ratios for the corresponding cores in our observational set. The results of these comparisons are shown in the plots of Figure 5 where the linear correlation coefficient for each plot is also given. From this analysis, although a linear correlation may be far too simplified, it is clear that there is a minor correlation between degree of anomaly and increasing density of the central source. How this correlation impacts the formation of the hyperfine anomalies is unclear and since in mappings of several of our cores in HCN J=3$\rightarrow$2, there appears to be a sustaining of the appearance of the anomalous profile where the density would be lower, it is clear that there are other physical factors contributing to the anomalous lineshape.

\thanks{The authors are thankful to Jonathan Rawlings, David Williams,
Tigran Khanzadyan, Paul Jones, Michael Burton, Indra Bains and Jonathan 
Tennyson for useful discussions. Sincere gratitude is given to Dr. Jungjoo 
Sohn for making available her HCN J=1$\rightarrow$0 data for 65 starless 
cores and to Dr. Shuro Takano for providing HCN data for the TMC-1 starless 
core. The use of the LAMDA database as well as the JCMT-OMP is acknowledged.}

\References


\refb Crapsi, A., Caselli, P., Walmsley, C. M., et al. 2005, ApJ,
619, 379

\refb Freed K. M., Mangum J. G., 2005, AAS, 37, \#184.08

\refb Gottlieb C. A., Lada C. J., Gottlieb E. W., Lilley A. E.,
Litvak M. M., 1975, ApJ, 202, 655

\refb Ladd N., Purcell C., Wong T., Robertson S., 2005, PASA,
22, 65

\refb Lo N., Cunningham M. R., Jones P. A., Bains I., Burton
M. G., Wong T., Muller E., Kramer C., Ossenkopf V.,
Henkel C., Deragopian G., Donnelly S., Ladd E. F., 2009,
MNRAS, 395, 1021

\refb Loughnane R. M., Redman M. P., Thompson M. A., Lo N., O' Dwyer B., 
Cunningham M., 2011 (submitted to MNRAS)

\refb Pirogov L., 1999, A\&A, 348, 600

\refb Papadopoulos, P. P., 2007, ApJ, 656, 792

\refb Snyder L. E., Buhl D., 1971, ApJ, 163, L47

\refb Sohn J., Lee C. W., Park Y. S., Lee H. M., Myers P. C.,
Lee Y., 2007, ApJ, 664, 928

\refb Tafalla M., Santiago-Garc\'{i}a J., Myers P. C., Caselli P.,
Walmsley C. M., Crapsi A., 2006, A\&A, 455, 577

\refb Walmsley C.M., Churchwell E., Nash A., Fitzpatrick E., 1982
ApJ, 258, L75 

\refb Wu Y., Zhu M., Wei Y., Xu D., Zhang Q., Fiege J. D.,
2005, ApJ, 628, 57

\end{document}